\documentclass[12pt]{article}
\usepackage{graphicx}

\begin{document}
\begin{center}
{ \large \bf 
Spin networks, quantum automata
and link invariants}
\end{center}

\vspace{24pt}

\noindent 
{\large {\sl Silvano Garnerone}}\\
\noindent Dipartimento di Fisica,
Politecnico di Torino,
corso Duca degli Abruzzi 24, 10129 Torino (Italy);\\ 
E-mail: silvano.garnerone@polito.it \\

\noindent
{\large {\sl Annalisa Marzuoli}}\\
\noindent Dipartimento di Fisica Nucleare e Teorica,
Universit\`a degli Studi di Pavia and 
Istituto Nazionale di Fisica Nucleare, Sezione di Pavia, 
via A. Bassi 6, 27100 Pavia (Italy);\\ 
E-mail: annalisa.marzuoli@pv.infn.it \\

\noindent
{\large {\sl Mario Rasetti}}\\
\noindent Dipartimento di Fisica and Istituto Nazionale di Fisica della Materia,\\
Politecnico di Torino,
corso Duca degli Abruzzi 24, 10129 Torino (Italy);\\ 
E-mail: mario.rasetti@polito.it \\

\begin{abstract}
The spin network simulator model represents a bridge between
(generalized) circuit schemes for standard quantum computation
and approaches based on notions from Topological 
Quantum Field Theories (TQFT). More precisely, when working 
with purely discrete unitary gates, 
the simulator is naturally modelled as families of 
quantum automata which in turn represent 
discrete versions of topological quantum computation models.
Such a quantum combinatorial scheme, which essentially encodes 
$SU(2)$ Racah--Wigner algebra and its braided counterpart, 
is particularly suitable 
to address problems in topology and group theory and we discuss
here a finite states--quantum automaton able to accept the language of
braid group in view of applications to the problem of  estimating 
link polynomials in Chern--Simons field theory.
\end{abstract} 

\vfill
\newpage

\section{Introduction}
In the past few years there has been a tumultuous activity 
aimed at introducing novel conceptual schemes for quantum 
computing. The approach proposed in \cite{MR1}, \cite{MR2}
relies on the (re)coupling theory of $SU(2)$ angular momenta
(Racah--Wigner tensor category) and
might be viewed as a generalization to 
arbitrary values of the spin variables of the usual quantum circuit
model based on qubits and Boolean gates.
Computational states belong to 
finite--dimensional Hilbert spaces labelled by both discrete 
and continuous parameters, and unitary gates may depend 
on quantum numbers ranging  over finite sets of values 
as well as continuous (angular) variables.
When working with purely discrete unitary gates, 
the computational space of the simulator is naturally modelled as families of 
quantum automata which in turn represent 
discrete versions of topological models of quantum computation
based on modular functors
of $SU(2)$  Chern--Simons theory ({\em cfr.} \cite{Fre} and references therein).
The discretized quantum theory underlying our scheme actually belongs to
the class of $SU(2)$ `state sum models' introduced in \cite{TV}
and widely used in $3$--dimensional simplicial quantum 
gravity ({\em cfr.} section 5 of \cite{MR2} and references therein).
From the computational viewpoint, we are in the presence of 
finite--states and discrete--time machines able to accept any (quantum)
language compatible with the Racah--Wigner algebra on the one hand, and as
powerful as Freedman's quantum field computer on the other.\\
As is getting more and more evident, the exponential efficiency 
that quantum algorithms may achieve {\em vs.}
classical ones might prove especially relevant in addressing problems
in which the space of solutions is not only endowed with a numerical (`digital')
representation but is itself characterized by some additional
`combinatorial' structure, definable in terms of 
the grammar and the syntax of some `language'.  
Our scheme can be easily adapted --by braiding the Racah--Wigner tensor category
as usually done in  $SU(2)$ Chern--Simons framework-- to make it accepting the language of
the braid group $B_n$ and to handle with (colored)
link polynomials expressed as expectation values of composite 
Wilson loop operators. 
In the last section we are going to exhibit a
quantum automaton calculation which processes efficiently --linearly in
the number of pairwise braidings-- the language of the braid group. This does not mean that we
have really got a `quantum algorithm' since we should show that
each `elementary' (in the sense of the spin network simulator) unitary gate
can be evaluated (or approximated) efficiently. This could be achieved by
exploiting both the combinatorial properties of the spin network graph
and suitable recursion relations which hold for  hypergeometric--type polynomials of discrete
variables (work is in progress in such directions).\\
Finally, let us point out
that the idea of using  braiding operators
to implement quantum gates actually dates back  to Freedman and collaborators \cite{Fre}
and has been exploited also by Kauffman and Lomonaco \cite{KL}. The common challenge
of all approaches
is, on the one hand,  the search for suitable  {\em unitary} representations of 
the braid group and, on the other, the selection of suitable encoding maps into quantum circuits
(or, eventually, into new quantum computing schemes).   
Aharonov, Jones and Landau have recently provided in \cite{AJL} an efficient quantum algorithm
 which approximates the Jones polynomial built from Markov traces
arising from representations of the braid group in the Temperly--Lieb algebra.

\section{Complexity of braids and links}

Let us recall some basic properties of the Artin braid group $B_n$.
 $B_n$ has $n$ generators, denoted by $\{\sigma_1,\sigma_2,\ldots,
\sigma_{n-1}\}$ plus the identity $e$, which satisfy the relations
$$
\sigma_i\,\sigma_j\,= \,\sigma_j\,\sigma_i \;\;\;\;\mbox{if}\;\,\,|i-j| > 1 \,\;\;(i,j=1,2,\ldots , n-1)
$$
\begin{equation}\label {braid}
\sigma_i\,\sigma_{i+1}\,\sigma_i\,= \,\sigma_{i+1}\,\sigma_{i}\,\sigma_{i+1}
 \;\;\;(\,i=1,2,\ldots ,n-2).
\end{equation}
An element of the braid group is a `word' $w$ in the standard generators of $B_n$,
{\em e.g.} $w= \sigma_3^{-1}\sigma_2$ $\sigma_3^{-1}\sigma_2 \, 
\sigma_1^{3}$ $\sigma_2^{-1}\sigma_1\sigma_2^{-2}$ $\in B_4$; the length $|w|$
of the word $w$ is the number of its `letters'.
The group acts naturally on topological sets of $n$ disjoint strands --
running downward and labeled from left to right --
in the sense that each generator $\sigma_i$ corresponds to the over--crossing of the
$i$th strand on the $(i+1)$th,
and $\sigma_i^{-1}$ represents the inverse operation (under--crossing)
according to $\sigma_i\,\sigma_i^{-1}$ $=\sigma_i^{-1}\sigma_i =e$.\par
\noindent As is well known, representation theory of Artin braid group enters heavily
into many physical applications, ranging from statistical mechanics 
to (topological) quantum field theories. Motivated by this remark,
the search for algorithms addressing computational problems arising in 
braid group context is becoming more and more compelling. 
Historically, three fundamental decision problems for any finitely presented group $G$
were formulated by Max Dehn in 1911:\par
\noindent $\bullet$ word problem: does there exist an algorithm to determine,
for any arbitrary word $w$ in the generators of $G$, whether or not $w=$ identity in $G$?\par
\noindent $\bullet$ coniugacy problem: does there exist an algorithm to decide
whether any pair of words in the generators of $G$ are conjugate to each other?\par 
\noindent $\bullet$ isomorphism problem: given an arbitrary pair of
finite presentations in some set of generators, does there exist an
algorithm to decide whether the groups they
present are isomorphic?
 
Following the development of the classical theory of
algorithms (recursive functions and Turing machine) it is reasonable to expect
that Dehn's problems might be recursively solvable or, at least, that
some `local' ones (the word and the coniugacy problems) be so. It turns out,
instead, that not only these problems, but a host of local and global decision
problems sharing a combinatorial flavor are unsolvable within such scheme.
As for the braid group, and referring to \cite{BB}
(which contains an exhaustive review together 
with an up--to--date bibliography), the word problem has been recently 
solved. More precisely, given two words $w,w'$ $\in B_n$
there exists an efficient algorithm  whose time
complexity function (namely the time required to perform the computation as
a function of the input `size') grows as $|w|^2 \, n \,log \, n$, where $|w|$ is the length of
the word $w$ and $n$ the index of the braid group. Thus this problem belongs
to the (classical) complexity class ${\bf P}$, which contains languages
accepted by a Turing machine in polynomial time. The word problem seems to be
one short step away from the non--minimal braid problem: given a word $w$ in
the generators (\ref{braid}) and their inverses, determine whether there is
a shorter word $w'$ in the same generators which represents the same element of
$B_n$. The decision process can be described as follows: list all the words that are shorter
than the given one, drop from the list as many candidates as possible by simple
criteria, and then test, one by one, whether the survivors represent the
same element as $w$. Surprisingly enough, this problem turns out to be ${\bf NP}$--complete,
namely belongs to the complexity class of {\bf N}on deterministic {\bf P}olynomial algorithm,
for which  a proposed solution can be checked efficiently (polynomially). 
${\bf NP}$--complete problems are in ${\bf NP}$ and can be characterized
by saying that a polynomial algorithm for one of them provides automatically
an efficient solution for all ${\bf NP}$ problems.
Finally, the best known classical algorithm for the coniugacy problem in 
the braid group is exponential in both $n$ and $|w|$.

Knot theory is closely related to braid group owing to Alexander's Theorem,
which states that every knot (or link ) $L$ in the $3$--sphere $S^3$ can be
represented (not uniquely) as a closed braid for some suitable $n$. Moreover,
the reduction from a planar diagram of the link $L$ to the closed braid $\hat{w}_{(n)} (L)$,
with $w_{(n)} \in B_n$, has been recently shown to be polynomial in the number
of strands (see the original references in \cite{BB}). The major challenge in knot
theory is of course the detection and classification of all possible knots and links
up to ambient isotopy.
There are a number of topological, combinatorial and algebraic invariants, starting from 
the group of the knot (the fundamental group of the complement of the knot
in $S^3$) and numerical invariants (linking number, crossing number, ..)
up to invariants of polynomial type. This last class of invariants is
related to the braid group, in the
sense that knot polynomials are  actually associated with characters (technically, Markov traces)
of (suitable) representations of $B_n$ for some $n$. 
A crucial role is played by the Jones polynomial \cite{VJ}, which, when evaluated
at particular roots of unity, turns out to be associated with suitable
expectation values of observables in 
Chern--Simons topological field theory \cite{Wi}. From the
computational side, it has been proved that the exact evaluation of the
polynomial $V_L (\omega)$ at $\omega =$ root of unity can be performed in
polynomial time in terms of the number of crossings of planar diagram of
$L$ if $\omega$ is a 2nd, 3rd, 4th, 6th root of unity. Otherwise, the problem is 
$\# {\bf P}$--hard \cite{JVW} (the computational complexity class 
$\# {\bf P}$--hard is the enumerative analog of ${\bf NP}$--complete problems).

Generally speaking, the exponential efficiency that quantum algorithms 
may achieve {\em vs.}
classical ones might prove especially relevant in addressing problems
in which the space of solutions is not only endowed with a numerical
representation but is itself characterized by some additional
`combinatorial' structure, definable in terms of
a grammar and a syntax and thus suitable to be encoded naturally in
the spin network computing framework (which will be addressed in the following).
There are a number of problems that are not
easily formulated in numerical terms
and that are quite often intractable in classical complexity theory
({\em cfr.} \cite{GJ}). In combinatorial and algebraic
topology typical issues are: the construction of presentations of the
fundamental group (or the first homology group) of compact $3$--manifolds
decomposed as handlebodies; the study of equivalence classes of knots/links in
the three--sphere, related in turn to the classification of hyperbolic
$3$--manifolds; the enumeration of inequivalent triangulations of
$D$--dimensional compact manifolds.\\ 
Of course efficient solutions of this kind
of problems would be interesting also in view of applications to
(discretized) models for quantum gravity at least in low spacetime dimensions.

\section{Spin network quantum simulator}

The theory of binary coupling of $N=n+1$ $SU(2)$ angular momenta
represents the generalization to an arbitrary $N$ of the coupling of two
angular momentum operators ${\bf J}_1, {\bf J}_2$
which involves Clebsch--Gordan (or Wigner) coefficients in their role
of unitary transformations between uncoupled and coupled basis vectors,
$|j_1\,m_1>\,\otimes\, |j_2\,m_2>$ and $|j_1\,j_2;JM>$ respectively.
The quantum numbers $j_1,j_2$ associated with ${\bf J}_1, {\bf J}_2$
label irreducible representations of $SU(2)$
ranging over $\{0,1/2,1,3/2,\ldots\}$; $m_1,m_2$ are the magnetic quantum numbers,
$-j_i \leq m_i \leq j_i$ in integer steps; $J$ is the spin quantum number
of the total angular momentum operator ${\bf J}= {\bf J}_1 + {\bf J}_2$ whose
magnetic quantum number is $M=m_1+m_2$, $-J\leq M\leq J$. Here units are chosen for which 
$\hbar =1$ and we refer to \cite{Russi} for a complete account on the
theory of angular momentum in quantum physics.
On the other hand, $SU(2)$ `recoupling' theory --which deals with relationships between
distinct binary coupling schemes of $N$ angular momentum operators-- 
is a generalization to any $N$ of the simplest case
of three operators ${\bf J}_1, {\bf J}_2, {\bf J}_3$ which calls into play unitary transformations
known as Racah coefficients or $6j$ symbols. A full fledged review on this
advanced topic in the general  framework of Racah--Wigner algebra
 can be found in \cite{BL9}.

 The architecture of the `spin network' simulator 
worked out in 
\cite{MR2} relies extensively on recoupling
theory and can be better summarized by resorting to
a combinatorial setting where
the computational space is modelled as an $SU(2)$--fiber space structure
over a discrete base space $V$
\begin{equation}\label{1}
(V,\,\mathbf{C}^{2J+1},\,SU(2)^J)_n
\end{equation}
which encodes all possible computational Hilbert spaces as well
as  unitary gates for any fixed number $N=n+1$ of incoming angular momenta.

$\bullet$ The base space $V\;\doteq\;\{v(\mathbf{b})\}$ represents the vertex set of a regular,
$3$--valent graph $\mathbf{G}_n(V, E)$ whose cardinality is $|V| = (2n)!/n!$.
There exists a one--to--one correspondence
\begin{equation}\label{2}
\{v(\mathbf{b})\}  \longleftrightarrow \{{\cal H}^J_n\,(\mathbf{b})\}
\end{equation}
between the vertices of $\mathbf{G}_n(V, E)$ and the computational Hilbert spaces 
of the simulator.

The label ${\mathbf{b}}$ above has the following meaning --on which we shall 
extensively return later on: for any given pair $(n,{\bf J})$, all binary coupling 
schemes of the $n+1$ angular momenta $\bigl \{ {\bf J}_{\ell} \bigr \}$, identified 
by the quantum numbers $j_1, \dots , j_{n+1}$ plus $k_1, \dots , k_{n-1}$ (corresponding 
to the $n-1$ intermediate angular momenta $\bigl \{ {\bf K}_{i} \bigr \}$) and by the 
brackets defining the binary couplings, provide the `alphabet' in which quantum 
information is encoded (the rules and constraints of bracketing are instead part of 
the `syntax' of the resulting coding language). The Hilbert spaces ${\cal H}^J_n\, 
(k_1,\dots , k_{n-1})$ thus generated, each $(2J+1)$-dimensional, are spanned by 
complete orthonormal sets of states with quantum number label set ${\mathbf{B}}$ 
such as, {\em e.g.} for $n=3$, $\bigl \{ \bigl ( \bigl (j_1 \bigl (j_2j_3 \bigr )_{k_1}
\bigr )_{k_2}j_4 \bigr )_j$ , $\bigl ( \bigl (j_1j_2 \bigr )_{k'_1} \bigl ( j_3j_4 
\bigr )_{k'_2} \bigr )_j \bigr \}$.

More precisely, for a given value of $n$, ${\cal H}^J_n(\mathbf{b})$ is the simultaneous
eigenspace of the squares of $2(n+1)$ Hermitean, mutually commuting angular
momentum operators
${\bf J}_1,\;{\bf J}_2,\;{\bf J}_3,\ldots,{\bf J}_{n+1}\,$ with fixed sum
${\bf J}_1\,+\,{\bf J}_2\,+\,{\bf J}_3\,+\ldots+{\bf J}_{n+1}\;=\;{\bf J}$, of
the intermediate angular momentum operators
${\bf K}_1,\,{\bf K}_2,\,{\bf K}_3,\,\ldots,\,{\bf K}_{n-1}$
and of the operator $J_z$ (the projection of the total angular momentum $\bf{J}$
along the quantization axis). The associated quantum numbers are 
$j_1, j_2,\ldots,j_{n+1}$ $;\,J;$ $ k_1,k_2,\ldots,$ $k_{n-1}$ and $M$, where $-J \leq M
\leq$ in integer steps. If
${\cal H}^{j_1}\otimes$ ${\cal H}^{j_2}\otimes\cdots$ $\otimes 
{\cal H}^{j_{n}}\otimes {\cal H}^{j_{n+1}}$
denotes the factorized Hilbert space, namely the $(n+1)$--fold tensor product 
of the individual eigenspaces of the $({\bf J}_{\ell})^2\,$'s, the operators 
${\bf K}_i$'s represent intermediate angular momenta generated, through Clebsch--Gordan series, 
whenever a pair of ${\bf J}_{\ell}$'s are coupled. As an example, by coupling
sequentially the ${\bf J_{\ell}}$'s according to the scheme
$(\cdots(({\bf J}_1+{\bf J}_2)+{\bf J}_3)+\cdots+{\bf J}_{n+1})$ $={\bf J}$ -- which generates
$({\bf J}_1+{\bf J}_2)={\bf K}_1$,
$({\bf K}_1+{\bf J}_3)={\bf K}_2$, and so on --
we should get a binary bracketing structure of the type
$(\cdots((({\cal H}^{j_1}\otimes{\cal H}^{j_2})_{k_1}$ $\otimes{\cal H}^{j_3})_{k_2}
\otimes$ $\cdots \otimes
{\cal H}^{j_{n+1}})_{k_{n-1}})_J$, where for completeness we add an overall  
bracket labelled by the quantum
number of the total angular momentum $J$. Note that, as far as $j_{\ell}$'s
 quantum numbers are involved, any value belonging to 
 $\{0,1/2,1,3/2,\ldots \}$ is allowed, while the ranges of the $k_i$'s are suitably 
 constrained by Clebsch--Gordan decompositions
 ({\em e.g.} if $({\bf J}_1+{\bf J}_2)={\bf K}_1$ $\Rightarrow$ $|j_1-j_2| \leq$
 $k_1 \leq j_1+j_2$).
We denote a binary coupled basis of $(n+1)$ angular
momenta in the $JM$--representation 
and the corresponding Hilbert space introduced in (\ref{2}) as
$$\{\,|\,[j_1,\,j_2,\,j_3,\ldots,j_{n+1}]^{\mathbf{b}}\, ;k_1^{\mathbf{b}\,},\,k_2^{\mathbf{b}\,}
,\ldots,k_{n-1}^{\mathbf{b}}\, ;\,JM\, \rangle,\;
-J\leq M\leq J \}$$
\begin{equation}\label{4}
=\;{\cal H}^{J}_{\,n}\;(\mathbf{b})\;\doteq\;\mbox{span}\;\{\;|\,\mathbf{b}\,;JM\,\rangle_n\,\}\;,
\end{equation}
where  the string inside $[j_1,\,j_2,\,j_3,\ldots,j_{n+1}]^{\mathbf{b}\,}$ 
 is not necessarily
an ordered one, $\mathbf{b}$ $\mathbf{B}$indicates the current binary bracketing structure and 
the $k_i$'s are uniquely associated with the chain of pairwise couplings selected by $\mathbf{b}$.

$\bullet$ For a given value of $J$
each ${\cal H}^J_n (\mathbf{b})$ has dimension $(2J + 1)$ over 
$\mathbf{C}$ and thus there exists one isomorphism
\begin{equation}\label{5}
{\cal H}^J_n (\mathbf{b})\;\;\; \cong _{\,\mathbf{b}}\;\;\; \mathbf{C}^{2J+1}
\end{equation}
for each admissible binary coupling scheme $\mathbf{b}$ of $(n + 1)$ incoming spins.
The vector space $\mathbf{C}^{2J+1}$ is naturally interpreted as the typical fiber attached to each vertex
$v(\mathbf{b}) \in V$ of the fiber space structure (\ref{1}) through the isomorphism (\ref{5}).
In other words, Hilbert spaces corresponding to 
different bracketing schemes, although isomorphic, are not identical since they actually 
correspond to (partially) different complete sets of physical observables, namely for instance
$\{{\bf J}^2_1,\,{\bf J}^2_2,\,{\bf J}^2_{12},\,{\bf J}^2_3,\,{\bf J}^2,\,J_z\}$ and 
$\{{\bf J}^2_1,\,{\bf J}^2_2,\,{\bf J}^2_3,\,{\bf J}^2_{23},\,{\bf J}^2,\,J_z\}$
respectively (in particular, ${\bf J}^2_{12}$ and ${\bf J}^2_{23}$ cannot be measured 
simultaneously). On the mathematical side this remark reflects the fact that the tensor 
product $\otimes$ is not an associative operation.

$\bullet$
For what concerns unitary operations acting on the computational
Hilbert spaces (\ref{4}), we examine first unitary transformations 
associated with recoupling 
coefficients ($3nj$ symbols) of $SU(2)$ ($j$--gates in the
present quantum computing context). As  shown in
\cite{BL9} any such coefficient can be 
splitted into `elementary' $j$--gates, namely Racah and phase transforms.
A Racah transform applied to a basis vector is defined formally as
${\cal R}\;:$ $| \dots (\,( a\,b)_d \,c)_f \dots;JM \rangle$ $\mapsto$
$|\dots( a\,(b\,c)_e\,)_f \dots;JM \rangle$, 
where Latin letters $a,b,c,\ldots$ are used here to denote generic, 
both incoming ($j_{\ell}\,$'s
in the previous notation) 
and intermediate ($k_i\,$'s) spin quantum numbers.
Its explicit expression reads
$$|(a\,(b\,c)_e\,)_f\,;M\rangle$$
\begin{equation}\label{7} 
=\sum_{d}\,(-1)^{a+b+c+f}\; [(2d+1)
(2e+1)]^{1/2}
\left\{ \begin{array}{ccc}
a & b & d\\
c & f & e
\end{array}\right\}\;|(\,(a\,b)_d \,c)_f \,;M\rangle,
\end{equation}
where there appears the $6j$ symbol of $SU(2)$ and $f$ plays the role
of the total angular momentum quantum number.  Note that, according to the
Wigner--Eckart theorem, the quantum number $M$ (as well as the angular
part of wave functions) is not altered by such 
transformations, and that the same happens with $3nj$ symbols. 
On the other hand, the effect of a phase transform amounts to introducing a
suitable phase whenever two spin labels are swapped
\begin{equation}\label{6}
| \dots ( a\,b)_c \dots;JM \rangle\; = \;(-1)^{a+b-c}
\,|\dots( b\,a)_c \dots;JM \rangle. 
\end{equation}
These unitary operations are combinatorially encoded into 
the edge set $E = \{e\}$ of the graph $\mathbf{G}_n(V, E)$: $E$ is just
the subset of the Cartesian
product $(V \times V )$ selected by the action of these elementary $j$--gates. More precisely, 
an (undirected) arc between two vertices $v(\mathbf{b})$ and $v(\mathbf{b}')$
\begin{equation}\label{8}
e\,(\mathbf{b},\mathbf{b}')\;\doteq \;(v(\mathbf{b}),\, v(\mathbf{b}')) 
\;\in \;(V \times V)
\end{equation}
exists if, and only if, the underlying Hilbert spaces are related to each other by 
an elementary unitary operation (\ref{7}) or (\ref{6}). 
Note also that elements in $E$ can be considered as mappings
$(V\,\times\,\mathbf{C}^{2J+1})_n$ $\longrightarrow$
$(V\,\times\,\mathbf{C}^{2J+1})_n$
\begin{equation}\label{9}
\;\;\;\;\;\;\;(v(\mathbf{b}),\,{\cal H}^J_n (\mathbf{b})\,)\, \mapsto\,
(v(\mathbf{b}'),\,{\cal H}^J_n (\mathbf{b}')\,)
\end{equation}
connecting each given decorated vertex to one of its nearest 
vertices and thus define a `transport 
prescription in the horizontal sections' belonging to the total space
$(V \times \mathbf{C}^{2J+1})_n$ of the fiber space (\ref{1}). 
The crucial feature that characterizes the graph $\mathbf{G}_n(V, E)$ arises from 
compatibility conditions satisfied by $6j$ symbols in (\ref{7}), {\em cfr.} \cite{Russi}.\\
The Racah (triangular) identity, the Biedenharn--Elliott (pentagon) identity
and the orthogonality conditions for $6j$ symbols 
ensure indeed that any simple path in $\mathbf{G}_n(V, E)$ with fixed endpoints can 
be freely deformed into any other, providing identical quantum transition amplitudes
at the kinematical level.

To complete the description of the structure 
$(V,\,\mathbf{C}^{2J+1},\,SU(2)^J)_n$ we should call into play $M$--gates which act
on the angular dependence of vectors in ${\cal H}^J_n (\mathbf{b})$ by rotating them.
The shorthand notation
$SU(2)^J$ employed in (\ref{1}) actually refers to the group of Wigner rotations,
which in turn can be interpreted as actions of the automorphism group
of the fiber $\mathbf{C}^{2J+1}$. Since rotations in the $JM$ representation do
not alter the binary bracketing structure of vectors in computational Hilbert spaces
we might interpret these actions as `transport prescriptions
along the fiber'. However, we are going to exploit here only computational 
degrees of freedom associated with $j$--gates, and thus we do not need to
explicitate the action of gates along the fiber. For this reason, we switch
to the notation $\mathbf{G}_n (V, E)\, \times\, \mathbf{C}^{2J+1}$, instead of
(\ref{1}), to represent
the computational space of the spin network simulator. 

Let us point out that our model of quantum simulator actually complies
with a variety of computing schemes, ranging from 
circuit--type models and
finite states--automata up to discretized versions of `topological' quantum
computation. Inside each of these classes we may also think of different encoding schemes
to deal with particular problems. 
In this respect, problems
from low dimensional topology, geometry, group theory and 
graph theory (rather than from number theory) turn out to be particularly suitable 
to be addressed in this `quantum--combinatorial' framework. 

Generally speaking, a computation is a collection of 
step--by--step transition rules (gates), namely
a family  of `elementary unitary  operations' 
and we assume that it takes
one unit of the intrinsic discrete time variable to perform anyone of them.
In the combinatorial setting described above
such prescriptions amount to select (families of) `directed paths' in
$\mathbf{G}_n (V, E)\, \times\, \mathbf{C}^{2J+1}$
all starting from the same input state and ending 
in an admissible output state. A single path in this family is associated with a particular
algorithm supported by the given program. 
By a directed path $\cal{P}$ with fixed endpoints
we mean a (time) ordered sequence 
\begin{equation}\label{12}
|\mathbf{v}_{\mbox{in}}\,\rangle_n\equiv
|\mathbf{v}_{0}\,\rangle_n\rightarrow
|\mathbf{v}_{1}\,\rangle_n\rightarrow\cdots\rightarrow
|\mathbf{v}_{s}\,\rangle_n\rightarrow\cdots\rightarrow
|\mathbf{v}_{L}\,\rangle_n\equiv
|\mathbf{v}_{\mbox{out}}\,\rangle_n\;,
\end{equation}
where we use the shorthand notation $|\mathbf{v}_{s}\rangle_n$ for computational states and
$s=0,1,2,\ldots ,L({\cal P})$ is the lexicographical labelling of the states along the  path. 
$L({\cal P})$ is the length of the path ${\cal P}$
and $L({\cal P}) \cdot \tau \doteq T$ is the time required to perform the process in terms of
the discrete time unit $\tau$. 

A computation consists in evaluating the expectation value of the unitary
operator $\mathbf{U}_{{\cal P}}$
associated with the path ${\cal P}$, namely
\begin{equation}\label{13}
\langle \mathbf{v}_{\mbox{out}}\,|\,\mathbf{U}_{\cal{P}}\,|\,
\mathbf{v}_{\mbox{in}}\,\rangle_n.
\end{equation}
By taking advantage of the possibility of decomposing 
$\mathbf{U}_{{\cal P}}$
 uniquely into an ordered sequence of elementary gates, (\ref{13}) becomes
\begin{equation}\label{14}
\langle \mathbf{v}_{\mbox{out}}\,|\,\mathbf{U}_{{\cal P}}\,|\,
\mathbf{v}_{\mbox{in}}\,\rangle_n\;=\;
\lfloor\,
\prod_{s=0}^{L-1}\,
\langle \mathbf{v}_{s+1}\,|\,{\cal U}_{s,s+1}\,|\,
\mathbf{v}_{s}\,\rangle_n\;\rfloor_{{\cal P}}
\end{equation}
with $L\equiv L({\cal P})$ for short. The symbol  
$\lfloor \; \rfloor_{{\cal P}}$ denotes the ordered 
product along the path ${\cal P}$ 
and each elementary operation 
is rewritten as  ${\cal U}_{s, s+1}$ $(s =0,1,2, \ldots L({\cal P}))$
to stress its `one--step'
character.

\section{Finite states--automata}

The theory of automata and formal languages addresses in a rigorous way the notions 
of computing machines and computational processes . 
If $A$ is an alphabeth, made of letters, digits or other symbols, and $A^*$ denotes
the set of all finite sequences of words over $A$, a language 
${\cal L}$ over $A$ is a subset of $A^*$. The empty word is $\epsilon$, 
the concatenation of two words $u$ and $v$ 
is simply denoted by $uv$. 
In the sixties Noam Chomsky introduced a four level--hierarchy describing 
formal languages according to their internal structure, namely 
regular languages, context--free languages, context--sensitive languages 
and recoursively enumerable languages (see {\em e.g.} \cite{HU}). 
The processing of each language is inherently related to  a particular computing model. 
Here we are interested in finite states--automata, 
the machines able to accept regular languages. 
A deterministic finite states--automaton (DFA) consists of a finite set of states 
$S$, an input alphabeth $A$, 
a transition function $F:S\times A\rightarrow S$, an initial state 
$s_{in}$ and a set of accepting states 
$S_{acc}\subset S$. The automaton starts in $s_{in}$ and reads an input word $w$ from left to right. 
At the $i-th$ step, if the automaton reads the word $w_i$, it updates its state to $s'=F(s,w_i)$, 
where $s$ is the state of the automaton reading $w_i$. The word has been accepted if the 
final state reached after reading $w$ is in $S_{acc}$.\\ 
In the case of non--deterministic finite states--automaton (NFA) 
the transition function is defined as a map from $S\times A$ in $P(S)$, where $P(S)$ is the power set of $S$. 
After reading a particular symbol the transition can lead to different states according to 
some assigned probability  distribution. 
If an NFA has $n$ internal states, for each symbol $a\in A$ 
there is an $n\times n$ transition matrix $M_a$ 
for which $(M_a)_{ij}=1$ if and only if the transition from the state $i$ to the state 
$j$ is allowed once the symbol $a$ has been read.\\
 From a general point of view,
 `quantum' finite states--automata (QFA) are obtained from their classical probabilistic counterparts by moving 
from the notion of (classical) probability associated with transitions to 
quantum probability amplitudes. Computation takes place inside 
a suitable Hilbert space through unitary matrices. 
Following \cite{MC},
the measure-once quantum finite--automaton (MOQFA)
is defined as a 5-tuple $M=(Q,\Sigma,\delta,{\bf q}_0,{\bf q}_f)$, where: $Q$ is a finite set of states; $\Sigma$ 
is a finite input alphabet with an end--marker symbol $\#$; $\delta:Q\times \Sigma\rightarrow Q$ is the 
transition function; $\delta({\bf q},\sigma,{\bf q}')$ 
is the probability amplitude for the transition from the state ${\bf q}$ to the state ${\bf q}'$ 
upon reading symbol $\sigma$; 
the state ${\bf q}_0$ is the initial configuration of the system, and ${\bf q}_f$ is 
the accepted final state. For all 
states and symbols the function $\delta$ must be unitary.  
At the end of the computational process the automaton measures its configuration: 
if it is in an accepted state then the input is accepted, otherwise is rejected. 
The configuration of the automaton 
is in general a superposition of states in the Hilbert space where the automaton 
lives. The transition function is 
represented by a set of unitary matrices $U_{\sigma}(\sigma\in\Sigma)$, where 
$U_{\sigma}$ represents the unitary 
transition of the automaton reading the symbol $\sigma$. The probability 
amplitude for the automaton $M$ to accept 
the string $w$ is given by
\begin{equation}\label{qaut}
f_M (w) = \left\langle {\bf q}_f \right| U_w \left| {\bf q}_0 \right\rangle , 
\end{equation}
where explicit form of $f_{M}(w)$ defines the language ${\cal L}$ accepted by the automaton $M$.

The spin network computational space $\mathbf{G}_n(V,E)\times \mathbf{C}^{2J+1}$
 (for a fixed $n$) naturally encodes the structures that define (families of) quantum automata. 
 Any finite set of binary coupled states belonging to
 the computational Hilbert spaces (\ref{4})
may represent states of some automaton, while combinations of the unitary
operations introduced in (\ref{7}) and (\ref{6})
acting on such states are actually transition functions and the amplitude
(\ref{qaut}) complies with expectation values introduced in (\ref{14}). 
The inherently step--by--step character of transition functions is related 
to the existence of an intrinsic discrete--time variable, denoted by $\tau$ 
in the previous section. 
All such  features make the spin network the ideal 
candidate for handling with finite--states (discrete--time) automata able 
to accept any `quantum language' compatible with the algebra of  $SU(2)$ angular momenta. 
In the next section we shall provide an explicit example for this class of quantum language.

\section{Quantum automaton calculation\\
 of link invariants}

As we shall recall below, (colored) link invariants can be obtained as expectation values
of Wilson--type operators in the $3D$ quantum $SU(2)$ Chern--Simons theory for each fixed
value of the coupling constant $k$ (related to the deformation parameter of the
universal enveloping algebra $U_{q}(sl(2))$, $q=$ root of unity) \cite{Wi}.\\
From a purely  algebraic viewpoint, the basic ingredient for building up such link invariants
is the `tensor structure' naturally associated with the representation ring
of the Lie algebra of any simple compact group.
In the case of $SU(2)$ this structure is provided by (tensor products of) Hilbert spaces
supporting irreducible representations together with unitary morphisms between them:
these objects are collected in the so--called Racah--Wigner tensor category 
introduced in section 3 \cite{BL9}.
In order to deal with (planar diagrams of) links we need also to specify the 
eigenvalues of the braiding matrix to be associated with the crossings of the links
and this is easily achieved by `braiding' the Racah--Wigner category. In the present
context, it is natural to take advantage of quantum group techniques in order to `split' 
phase transforms like that in (\ref{6}) by assigning different wheights --depending
on the deformation parameter $q$-- to right and
left handed twists. From the combinatorial viewpoint, this generalization corresponds to
replace the spin network computational space 
$\mathbf{G}_n (V, E)\, \times\, \mathbf{C}^{2J+1}$ (encoding the Racah--Wigner category)
with its $q$--braided counterpart $(\mathbf{G}_n (V, E)\, \times\, \mathbf{C}^{2J+1})\,\times \,
\mathbf{Z}_2$, where $6j$ symbols in the Racah trasforms (\ref{7}) 
become  $q$--deformed.\\
As a general remark, let us point out that the link invariants we are going to address
are `universal' in the sense that historically distinct approaches ( R--matrix representations
obtained with the quantum group method, monodromy representations of the braid group
in $2D$ conformal field theories, the quasi tensor category approach by Drinfeld and the
$3D$  quantum Chern--Simons theory, {\em cfr.} \cite{Gua} for a review) are indeed different
aspects of the same underlying algebraic structure. In this sense the Chern--Simons setting
describes the universal structure of braid group representations shared by all
the models quoted above, and in particular the `universal' (colored) link polynomial 
essentially coincides
with the Reshetikhin--Turaev invariant \cite{RT}. However, when the group is $SU(2)$, 
we may speak of `extended' Jones polynomials since the Jones polynomial \cite{VJ} is recovered 
by selecting the $j=\frac{1}{2}$ representation on each of the link component
(or on each strand of the associated braid). On the other hand, the topological quantum
field approach calls into play links embedded in three--dimensional manifolds, and thus,
for instance, the invariants can be naturally interpreted as quantum invariants  
of hyperbolic manifolds (complements of links in the $3$-sphere).

Recall that the $SU(2)$ Chern--Simons action for the $3$--sphere $S^3$ is
\begin{equation}\label{CSaction}
k\,S(A)=\frac{k}{4\pi}\int_{S^3}tr(AdA+\frac{2}{3}A \wedge A \wedge A)
\end{equation}
where $A$ is the connection one--form with value in the Lie algebra of the 
gauge group $SU(2)$ and $k$ is the coupling constant (constrained to be 
a positive integer by the gauge--invariant quantization procedure). 
The observables of the associated quantum field theory 
are Wilson loop operators defined, for an oriented knot 
$K$ carrying a spin--$j$ representation, as
\begin{equation}\label{WilK}
W_j\,[K]=tr_j\,Pexp\oint_KA
\end{equation}
and, for a link $L$ made of $s$ components $K_l$, each labelled by a 
spin (coloring),
\begin{equation}\label{WilL}
W_{j_1j_2\ldots j_s}\,[L]\,=\,\prod_{l=1}^s\;W_{j_l}\,[K_l].
\end{equation}
The expectation values of the above operators are formally given by
\begin{equation}\label{expect}
V_{j_1...j_s}\,[L]\,=\,\frac{\int_{S^3}\,[dA]\,W_{j_1...j_s}\,
[L]\,e^{ikS(A)}}{\int_{S^3}\,[dA]\,e^{ikS(A)}}.
\end{equation}
where  the functional integration is
over flat $SU(2)$--connections on $S^3$ and
the generating functional in the denominator will be normalized to $1$.
Such expectation values
depend on the isotopy type of the oriented 
link $L$ and on the set of representations labelled by $(j_1,...,j_s)$ associated 
with the link components (and do not depend on the metric
of the ambient manifold). To evaluate explicitly such functional averages we have to go through
the relationship between Chern--Simons theory on an oriented, compact three--manifold 
with boundary and the  Wess--Zumino--Witten (WZW) conformal field theory which lives
on the $2$--dimensional boundary components. 
If $\Sigma^{(1)},\Sigma^{(2)},\ldots,\Sigma^{(r)}$ are the boundary components,
(a finite number of) Wilson 
lines (carrying labels $j_l$) intersect $ \Sigma^{(i)}$ at some 
points (punctures) $P_l^{(i)}$.  According to the axioms for TQFT, we associates with 
each boundary component $ \Sigma^{(i)}$ a finite--dimensional  Hilbert space 
denoted by $H^{(i)}$.
In this case the Chern--Simons functional (represented geometrically  by a cobordism between incoming and 
outgoing boundaries and algebraically by a functor between the associated Hilbert spaces) 
is a state belonging to 
the tensor product of the boundary Hilbert spaces. It can be shown that the conformal blocks 
of the $SU(2)_{\ell}$ WZW field theory on the boundaries with 
punctures actually provide basis vectors for the Hilbert spaces $H^{(i)}$
(the level $\ell$ of the WZW model is related to the deformation
parameter $q$ according to $q = \exp \{- 2\pi i/(\ell + 2)\}$, and in turn $\ell$
is related to the coupling constant $k (\geq 3)$ of the CS theory in the bulk by $\ell=k-2$).\\
Suppose that the oriented link appearing in the
the expectation value (\ref{expect}) is endowed with a plat presentation \cite{Bir}, namely
is actually the closure of an oriented braid with $2m$ strands.
If we remove two open three--balls from $S^3$ we get two boundaries (both topologically $S^2$,
but with opposite orientations) and
we can always accomodate $2m$ 
`unbraided' Wilson lines carrying labels $j_1,j_2,...,j_{2m}$
starting from the incoming (lower) boundary and ending into the outgoing (upper) one.
Following the formulation due to Kaul \cite{Kaul}, let us start by denoting 
this `identity' oriented  colored oriented braid as
\begin{equation}\label{idbraid}
\nu_{\,I}\;\left( \begin{array}{cccc}
   \widehat{j}_1^*  & \widehat{j}_2^*  & \ldots & \widehat{j}_{2m}^*   \\
   \widehat{j}_1  & \widehat{j}_2  & \ldots & \widehat{j}_{2m}   
\end{array} \right),
\end{equation}
where $\widehat{j}_i =(j_i, \epsilon_i)$ $i=1,2,\ldots 2m$ represents the spin $j_i$ together with
an orientation $\epsilon_i, i=\pm 1$ for a strand going into or away from the boundary,
while stars over the symbols stand for the opposite choice of the
orientation. A generic colored oriented braid on $2m$ strand can be generated from 
this identity braid by
applying a suitable `braiding' operator $B$ (written in terms of generators $B_1,B_2,...,B_{2m-1}$) 
starting from the lower boundary. The resulting colored oriented braid is denoted by
\begin{equation}\label{genbraid}
\nu _{\,B}\; \left( {\begin{array}{ccccc}
   {\widehat{j}_1 } & {\widehat{j}_1^* } & {\ldots} & {\widehat{j}_m } & {\widehat{j}_m^* }  \\
   {\widehat{l}_1 } & {\widehat{l}_1^* } & {\ldots} & {\widehat{l}_m } & {\widehat{l}_m^* }  \\
 \end{array} } \right) , 
\end{equation}
where labels have been ordered according to the requirement of having a plat
presentation for the associated link. 
According to our previous remark, we must now select bases in the two boundary Hilbert
spaces $H^{(1)}$ and $H^{(2)}$ and then built up the Chern--Simons functional
which will belong to $H^{(1)} \otimes H^{(2)}$.
There are basically two possible choices of such bases on the incoming Hilbert space
$H^{(1)}$,
corresponding to two sets of correlators  in WZW theory involving $2m$ 
primary fields with angular momentum assignments $(j_1,\ldots,j_{2m})$, namely
\begin{equation}\label{WZWvect}
|\;( (j_1 \, j_2) \ldots (j_{2m - 1} \, j_{2m}) )_{\,[p;r]}^{\,0}\; \rangle ;\;\;
|\;( ( j_1 \,( j_2 j_3 )\ldots
( j_{2m - 2} j_{2m - 1}))\,j_{2m})_{\,[t;s]}^{\,0}  \;\rangle.
\end{equation}
The shorthand notations $[p;r]$ and $[t;s]$ represent the strings of intermediate
spin variables arising from the given pairwise couplings of the incoming $(j_1,\ldots,j_{2m})$, 
$\{p_0,p_1,\ldots ;r_0,r_1,\ldots\}$ and 
$\{t_0,t_1,\ldots; s_0,s_1,\ldots\}$ respectively. Moreover, the superscript $^0$
corresponds to the requirement that all such pairings must generate 
$SU(2)$ singlets (spin--$0$ representations). Unitary transformations between the two sets of
vectors (\ref{WZWvect}) are implemented by the so--called
duality matrices, involving $q$--deformed $6j$ symbols ({\em cfr.} the appendix of \cite{Kaul}
for their explicit expressions).\\
When the outgoing Hilbert space is considered (corresponding to the boundary
$\Sigma^{(2)}$ endowed with the opposite orientation with respect to $\Sigma^{(1)}$)
we have to introduce bases which are dual with respect to (\ref{WZWvect}), 
suitably normalized according to 
\begin{equation}\label{dualba1}
\langle \;(( {j_1 \,j_2 } ) \ldots( j_{2m - 1}\, j_{2m}))_{\,[p;r]}^{\,0}\;|
\;( (j_1 \, j_2) \ldots (j_{2m - 1} \, j_{2m}) )_{\,[p';r']}^{\,0}\; \rangle\;=\;
\delta_{(p)(p')}\: \delta_{(r)(r')}
\end{equation}
and
\begin{equation}\label{dualba2}
\langle (( j_1 ( j_2 j_3 )
( j_{2m - 2} j_{2m - 1}))j_{2m})_{\,[t;s]}^{\,0}\;  |\,
( ( j_1 \,( j_2 j_3 )\ldots
( j_{2m - 2} j_{2m - 1}))\,j_{2m})_{\,[t';s']}^{\,0}  \;\rangle =
\delta_{(t)(t')}\, \delta_{(s)(s')}.
\end{equation}
The basis vectors (\ref{WZWvect})
are eigenfunctions of the odd and even braid generators, respectively
$$
B_{2l + 1} | (( j_1 j_2 )...( j_{2l + 1} j_{2l + 2} )... )_{[p;r]}^0 \rangle  
\,=\,
\lambda _{p_l} ( \hat{j}_{2l + 1} ,\hat{j}_{2l + 2}  )
| (( {j_1 j_2 })...( j_{2l + 2} j_{2l + 1}  )... )_{[p;r]}^0  \rangle 
$$
\begin{equation}\label{brop}
B_{2l} | ( ( j_{2m} j_1  )...( j_{2l} j_{2l + 1}  )... )_{[t;s]}^0 \rangle \, =\, 
\lambda _{t_l } 
( \hat{j}_{2l} ,\hat{j}_{2l + 1}  )| ( ( j_{2m} j_1  )
...( j_{2l + 1} j_{2l}  )...)_{[t;s]}^0  \rangle 
\end{equation}
with eigenvalues given by the expressions
$$
 \lambda_z \,( \hat{j},\hat{j'}) = \lambda _z^{(+)} ( j,j') 
= (-)^{j + j' - z}\, q^{(c_j  + c_{j'})/2 
+ c_{\min (j,j')} - c_z /2} \;\; \mbox{if} \;\;\epsilon \epsilon'=+1
$$
\begin{equation}\label{breigen}  
\lambda_z \,( \hat{j},\hat{j'}) = ( \lambda _z^{(-)} ( {j,j'}) )^{ - 1}  
=(-)^{|j - j'| - z} \, q^{|(c_j  - c_{j'}|/2 - c_z /2} \;\; \mbox{if} \;\;\epsilon \epsilon'=-1
\end{equation}
where $q$ is the deformation parameter, $z \in \{p_l,t_l\}$ and $c_z  \equiv z(z + 1)$ 
is the quadratic Casimir for the spin $z$ representation. Thus 
$
\lambda _z^{(+)}(j,j')
$ 
is the eigenvalue of the matrix which performs a right handed 
half--twist in strands with the same orientation, while 
$
\lambda _z^{(-)}(j,j')
$ 
is the eigenvalue of the matrix which performs a right handed 
half--twist in strands with the opposite orientations. 
The expectation value of 
the Wilson operator (\ref{expect}) for any link $L$ presented as the plat closure
of an oriented braid over $2m$ strands can be recasted into the expectation
value of the braiding operator $B$ associated with the oriented braid
(\ref{genbraid}) according to
$$
V_{j_1 j_1^* \ldots j_m j_m^*} \;\left[ L \right] = \left( {\prod\limits_{i = 1}^m {\left[ {2j_i  + 1} 
\right]} } \right)\times
$$
\begin{equation}\label{poly}
\left\langle {\left( {\left( {\widehat{l}_1 \widehat{l}_1^* } \right)...
\left( {\widehat{l}_m \widehat{l}_m^* } \right)} \right)_{\left[ {0;0} 
\right]}^0 } \right|\,B\left( {\begin{array}{ccccc}
   {\widehat{j}_1 } & {\widehat{j}_1^* } & {...} & {\widehat{j}_m } & {\widehat{j}_m^* }  \\
   {\widehat{l}_1 } & {\widehat{l}_1^* } & {...} & {\widehat{l}_m } & {\widehat{l}_m^* }  \\
\end{array} } \right)\left| {\left( {\left( {\widehat{j}_1 \widehat{j}_1^* } 
\right)...\left( {\widehat{j}_m \widehat{j}_m^* } \right)} \right)_{\left[ {0;0} 
\right]}^0 } \right\rangle 
\end{equation}
This expression (which is normalized according to the standard conventions) gives the value
of the extended Jones polynomial of the link $L$. The operator $B$ is expressed 
in terms of (a finite sequence of)  braid generators (\ref{brop}), suitably
converted in terms of the current basis vectors by acting with $q$--Racah transforms
on the even subset, while $[2j_i+1]$
is the $q$--dimension of the representation $j_i$.

Coming back to the definition of the quantum finite state--automaton (QFA)
given in the previous section, namely a
5-tuple $M=(Q,\Sigma,\delta,{\bf q}_0,F)$, we recognize that, given a plat presentation 
of a link $L$, we can always build up a QFA $M_L$ which recognizes the 
language of the braid group and whose associated
probability amplitude (\ref{qaut}) is given by the value of the extended
Jones polynomial (\ref{poly}). More precisely: ${\bf q}_0$ is a particular ket vector of the type
(\ref{WZWvect}); $\Sigma$ is the alphabeth given by the $(2m-1)$ braiding
operators (\ref{brop}); $F$ is a set of final stated suitably selected among
the bra vectors in (\ref{WZWvect}). Note that, according to the expression given in
(\ref{poly}), $F$ must contain (singlet) states which possess the same pairwise coupling structure 
as the initial state ${\bf q}_0$ exhibits. Owing to the topological properties of
the plat presentation, this means that final states may differ from 
${\bf q}_0$ by a permutation on the string $(j_1 j_2 \ldots j_m)$
and thus we can actually build up a family of $m!$ automata out of one
initial state ${\bf q}_0$. Once a final state ${\bf q}_{f}$ has been 
selected (the right permutation
can be singled out in a fast way even by a classical machine) the evaluation of
({\ref{poly}) is carried out by the automaton linearly in the length of the 
`word' $B$. The length of $B$ is the number of $\{$ braiding operators
+ $q$--Racah transforms $\}$ entering into the explicit  expression of
$B$ and equals the number of crossings of the plat presentation of $L$. 
It can be easily recognized that such an automaton can be associated with a 
particular path in the $q$--braided spin network computational space 
$(\mathbf{G}_n (V, E)\, \times\, \mathbf{C}^{2J+1})\,\times \,
\mathbf{Z}_2$ for $n=2m-1$ since the process carried out here is
a particular realization of the general computing procedure described at the
end of section 3 ({\em cfr.} (\ref{14})).\\
What we have really done is to `encode' the combinatorial structure underlying quantum $SU(2)$
Chern--Simons theory (and the associated  WZW boundary theory)
at some fixed level $\ell$ into the above
abstract $q$--braided $SU(2)$--decorated graph. This does not mean, of course, 
that we have got a
quantum algorithm in the proper sense, since the encoding
map could not be `efficient'  (nor efficiently approximated)
with respect to standard models of computation
(Boolean circuit, Turing machine). We are currently addressing such open problems.

\end{document}